# Analysis, Modification, and Implementation(AMI) of Scheduling Algorithm for the IEEE 802.116e (Mobile WiMAX)

C. Ravichandiran[1], Dr. C. Pethuru Raj[2], Dr. V. Vaidhyanathan[3],

[1] IT Leader, Zagro Singapore Pte Ltd, Singapore.
[2] Lead Architect, CR Division of Robert Bosch, Bangalore, India.
[3] Professor and HOD, Dept. of IT, SASTRA University, India.

*Abstract*— Mobile WiMAX (Worldwide Interoperability for Microwave Access) is being touted as the most promising and potential broadband wireless technology. And the popularity rate has been surging to newer heights as the knowledge-backed service era unfolds steadily. Especially Mobile WiMAX is being projected as a real and strategic boon for developing counties such as India due to its wireless coverage acreage is phenomenally high. Mobile WiMAX has spurred tremendous interest from operators seeking to deploy high-performance yet cost-effective broadband wireless networks. The IEEE 802.16e standard based Mobile WiMAX system will be investigated for the purpose of Quality of Service provisioning. As a technical challenge, radio resource management will be primarily considered and main is the costly spectrum and the increasingly more demanding applications with ever growing number of subscribers. It is necessary to provide Quality of Service (QoS) guaranteed with different characteristics. As a possible solution the scheduling algorithms will be taken into main consideration and the present well known algorithms will be described.

In this paper, we have highlighted the following critical issues for Mobile WiMAX technologies. This paper specifically discussed about the below mentioned in detail.

- QoS Requirements For IEEE 802.16 Service Classes, Achieving efficient radio resource management

- Deficit Round Robin (DRR) Scheduling algorithm

- Modified Deficit Round Robin (MDRR) scheduling algorithm's attributes, properties and architecture

- System Model And Scenarios Using OPNET Modeler Software

- Simulation Limitations And Constraints

I. INTRODUCTION

"Mobile WiMAX" refers to a rapidly growing broadband wireless access solution built upon the IEEE 802.16e-2005 air interface standard. It is equally applicable to fixed, portable and mobile applications. The Mobile WiMAX air interface utilizes Orthogonal Frequency Division Multiple Access (OFDMA) for improved multipath performance in non-line-of-sight (NLOS) environments and high flexibility in allocating resources to users with different data rate requirements. The fundamental premise of the IEEE 802.16e media access control (MAC) architecture is QoS. Mobile WiMAX QoS features enable operators to optimize network performance depending on the service type (e.g., voice, video, and gaming) and the user's service level. In wireless communication the task of taking care of resources being utilized falls into Radio Resource Management (RRM). RRM, in general, is responsible for the improvement of efficiency and reliability of radio links, but in particular it enables many specific operations are below:

Rate control: To be capable of utilizing the bandwidth more efficiently and maintain the quality of the radio links Adaptive Modulation and Coding (AMC) technique is used in wireless communication. Channel assignment: Mapping the most efficient subcarriers to their corresponding symbol times is done with the help of the information provided trough RRM Subcarrier permutation: Mainly there are two types of subcarrier permutation: Distributed Subcarrier Permutation: where frequency subcarriers are spread along the whole allocated transmission band pseudo-randomly.

Scheduling System: Scheduling makes up an important part of the communication systems since it is chiefly the process of sharing the bandwidth. Therefore, has a significant effect on: The time taken by a packet to travel from one OSI (Open System Interconnection) stack layer to its corresponding peer layer. Jitter: The inter-packet arrival time difference. Packet Loss: The amount of packet being dropped on both the Uplink (UL) and Downlink (DL). Throughput: The number of successful bits/packets per second arriving at the receiver [1].

II. IEEE 802.16E ( MOBILE WIMAX)

Mobile WiMAX is expected to deliver significant improvements over Fixed WiMAX which makes it even more attractive for fixed deployments. In wireless environments, link budget (measured in dB) and spectral efficiency are the two primary parameters used for evaluating system performance [22]. Mobile WiMAX) standard amendment as the investigation host for the discussed scheduling algorithms, thus from now on any mentioned technology will be those which are valid for the Mobile WiMAX. For instance, the multiple access technique considered will be Orthogonal Frequency Division Multiple Access (OFDMA) rather than the previously utilized Orthogonal Frequency Division Multiplexing (OFDM) technique.

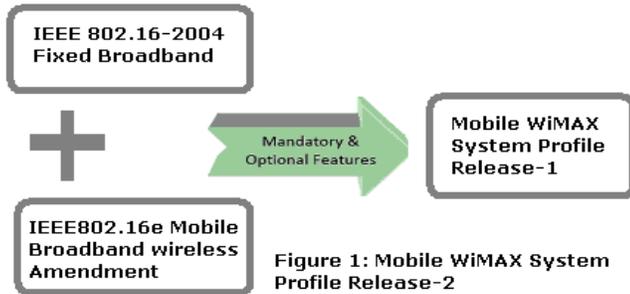

Figure 1: Mobile WiMAX System Profile Release-2

Since the packet scheduling theme is mainly done in the MAC layer of the OSI protocol stack, the main task will be to present a brief background overview on the scheduling procedure related MAC entities; as it will also briefly explain the WiMAX Physical (PHY) Layer inter related functionalities which are involved in the scheduling process[2].

### A. IEEE 802.16E PHY Layer

The WiMAX physical layer is based on orthogonal frequency division multiplexing. OFDM is the transmission scheme of choice to enable high-speed data, Big files, video, Deficit Round Robin (DRR), and multimedia file and is used by a variety of commercial broadband systems, including DSL, Wireless, Digital Video Broadcast-Handheld (DVB-H), and MediaFLO, besides WiMAX[22].

Basically, OFDM subdivides the channel to multiple subcarriers, where each subcarrier is orthogonal to the other. There are three types of subcarriers:

- Null Subcarrier: used as guard band or DC carriers.
- Pilot Subcarrier: used for channel estimation and channel tracking.
- Data Subcarrier: carrying the data information.

Figure 2 illustrates the types of the subcarriers in a 10 MHz channel bandwidth.

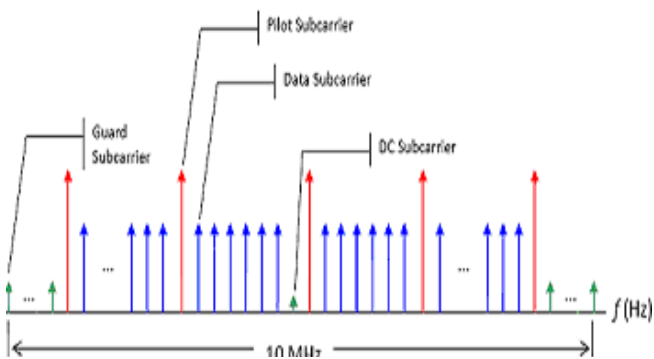

Figure 2: OFDM Sub carrier Structure

Fundamentally, OFDMA is OFDM with the application of Subchannelization and Time Division Multiple Access (TDMA). Subchannelization basically means to group multiple subcarriers and allocate them to a single user over one, two or thee OFDMA symbol time(s). Figure 2 could be modified to show Subchannelization for OFDMA, as illustrated in Figure 3:

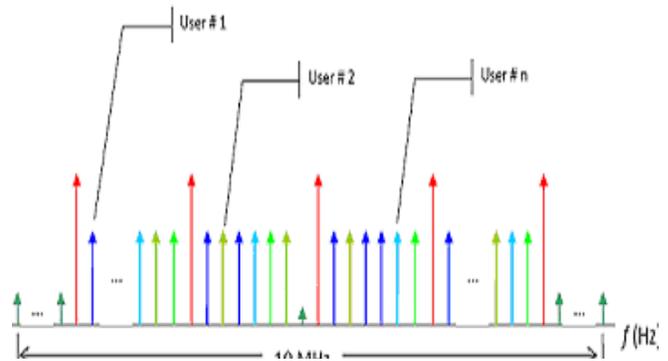

Figure3: Subchannelization in OFDM

Note that different colours mean different users. Unlike OFDM, here in OFDMA data streams from different users are multiplexed, rather than using the whole band for a single user per one symbol frame. Note also that the subcarriers are not adjacently grouped, but in a random manner. This introduces frequency diversity which is specifically rewarding in the case of mobile communications (since the channel tends to vary the most among other cases). Adding to that, it allows a better application of fairness between the users since the probability of a user experiencing bad channel impulse response will be less.

### B. IEEE 802.16e MAC Layer

IEEE 802.16 MAC was designed for point-to-multipoint broadband wireless access applications. The primary task of the WiMAX MAC layer is to provide an interface between the higher transport layers and the physical layer. The 802.16 MAC is based on collision sense multiple access with collision avoidance (CSMA/CA). The MAC incorporates several features suitable for a broad range of applications at different mobility rates, as mentioned below[21][7]:

- Broadcast and multicast support.
- Manageability primitives.
- High-speed handover and mobility management primitives.
- Three power management levels, normal operation, sleep and idle.
- Header suppression, packing and fragmentation for efficient use of spectrum.
- Five service classes, unsolicited grant service (UGS), real-time polling service (rtPS), non-real-time polling service (nrtPS), best effort (BE) and Extended real-time variable rate (ERT-VR) service.

### III. SCHEDULING ALGORITHMS: ANAYSIS AND MODIFICATION

### A. Scheduling Algorithms

Packet Switching (PS) networks came into existence, need was recognized to differentiate between different types of packets. Since then packet scheduling has been a hot research subject and its still being investigated at many institutions/company. This is basically because scheduling means bandwidth sharing.

Traditionally, the First Come First Served (FCFS) scheme had been used for packet scheduling. Packets coming from all the input links were enqueued into a First In First Out (FIFO) memory stack, then they were dequeued one by one on to the output link. This is shown in Figure 4-(a). Since unlike packets were mixed and treated equally, packets requiring urgent delivery could not be achieved. So there is no scheduling action taking place in this case.

In the present time different queues are specified to non similar packets for achieving packet classification. In this case scheduling should be done. The main task of the embedded scheduling algorithm is to choose the next packet to be dequeued from the available multi queues and forwarded onto the output link. This is illustrated in Figure 4- (b) shown below.

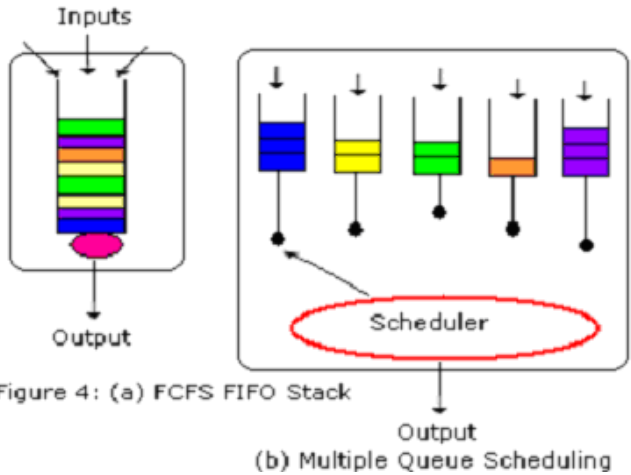

Figure 4: (a) FCFS FIFO Stack
(b) Multiple Queue Scheduling

*B. Scheduling Algorithm Aim*

The main aims behind the concept of packet scheduling could be simply defined by four points:

- The first and foremost goal of a scheduling algorithm is to be able to share the total system bandwidth fairly.
- The algorithm should be able to guarantee the minimum bandwidth per SS. This consequently leads to the separation of different demanding SSs.
- To be capable of meeting packet drop guarantees.
- To be able to assure latency guarantees.
- To be capable of reducing latency variation.

*C. Scheduling Algorithm Criterions*

On the other hand, metrics on which scheduling algorithms will be compared are as follows:

**Simplicity:** This criterion in particular is of significant importance to the high-speed networks available in the present time since a simply implemented algorithm directly leads to a faster operation and thus lower packet latency. In addition, a less complex algorithm may also mean a lower implementation cost. Furthermore, simplicity of the algorithm would also benefit mobile devices that possess a limited power resource.

**Fairness:** Fairness can be defined such that: "If two flows are backlogged, difference between their weighted throughputs is bounded." Since mobile subscriber stations are considered equal regardless their location, distance from the BS and channel quality, a scheduling algorithm should be able to recompense users with poor channel quality and it is based on the "Max-min fair share".

**Flexibility:** A scheduling algorithm should be capable of accommodating users with different QoS requirements.

**Link Utilization:** Maximizing the link utilization, especially in the case of wireless communications, is of great significance to service providers. Since the produced income is directly proportional to this criteria.

**Protection:** A scheduling algorithm is required to protect well-behaving users from the misbehaving users. Well-behaving users are those who stick to the Service Level Agreements (SLA), while the misbehaving users are those who do not comply with the SLA at all times and thus causes unpredictability in the network.

IV.  FUNDAMENTAL SCHEDULING ALGORITHM

In the coming subsections the fundamental scheduling algorithms will be briefly described. These basic algorithms make up the foundation of target scheduling algorithm, the Modified Deficit Round Robin (MDRR). Afterwards, a detailed investigation of MDRR will be carried out, emphasizing the modifications made to adjust the algorithm.

*A. Round Robin(RR)*

Round-Robin as a scheduling algorithm is considered the most basic and the least complex scheduling algorithm. Basically the algorithm services the backlogged queues in a round robin fashion. Each time the scheduler pointer stop at a particular queue, one packet is dequeued from that queue and then the scheduler pointer goes to the next queue. This is shown in Figure 5.

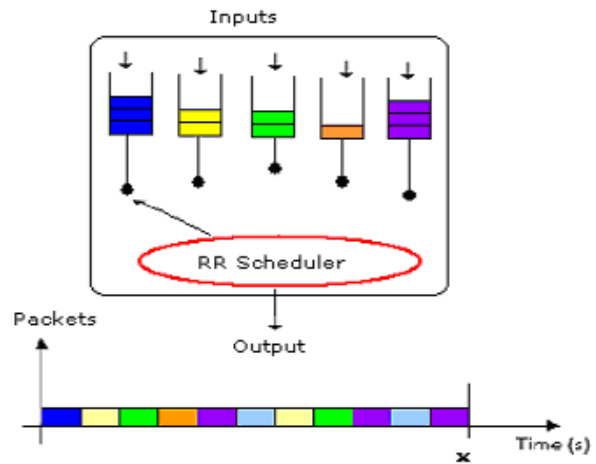

Figure 5: Round Robin Scheduler

Note that in this case all packets are of same length. However, for instance an MPEG video application may have variable size packet lengths. This case is shown in Figure 6.

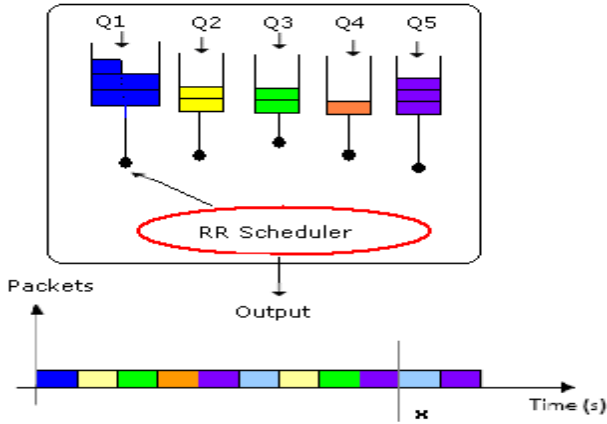

Figure 6: Round Robin Scheduler - Variable Packet size

It is assumed that queues Q2-Q5 have constant packet size of 50 bytes and Q1 have a packet size of 100 bytes. Note that in Figure 6, unlike Figure 5, Q1 has superior throughput than the other queues.

- Previously Q1 was transmitting 3x50 bytes per X interval = 150 bytes/X interval
- Now Q1 is transmitting 2x100 bytes per X interval = 200 bytes/X interval

This was caused by transmitting longer packet lengths. Hence, we can deduce that the round robin scheduling algorithm does not convey fairness in systems with variable packet lengths, since RR tends to serve flows with longer packets more[4].

### B. Weighted Round Robin(WRR)

Weighted round robin was designed to differentiate flows or queues to enable various service rates. It operates on the same bases of RR scheduling. However, unlike RR, WRR assigns a weight to each queue. The weight of an individual queue is equal to the relative share of the available system bandwidth. This means that, the number of packets dequeued from a queue varies according to the weight assigned to that queue. Consequently, this differentiation enables prioritization among the queues, and thus the SSes. Nevertheless, the downside of a WRR scheduler, just like an RR scheduler is that, different packet lengths being used by SSes would lead to the loss of its fairness criterion.

### C. Deficit Round Robin (DRR)

The DRR scheduling algorithm was designed to overcome the unfairness characteristic of the previous RR and WRR algorithms. In DRR scheduling, every queue is accompanied by a deficit counter which is initially set to the quantum of the queue. A quantum is a configurable amount of credits (in bits/bytes) given to a queue whenever it is served. Quantum should represent the idle amount of bits/bytes a queue may require. Adding quantum is proportional to assigning weight to a queue.

The deficit counter is increased by one quantum on every visit of the scheduler, except when the queue is empty; the deficit counter is deducted by the amount of information being served on each pass of the scheduler to the queue. Queues are served only if the amount of quantum added to the remaining deficit counter amount from previous round is greater tan zero. Otherwise, the quantum is added only and that particular queue is held till it is served in the next round.

On the other hand, when packets of a backlog queue are completely served then any remaining credit in the deficit counter will be set to zero, as the accumulation of credit without being utilized will result in unfairness [3].

```
// qI = Quantum for Queue i;
// dcI = Deficit Counter for Queue i;
// n = Maximum Packet Size;
// pI = Packets in Queue i;
// noP = Number of Packets;
// pS = Packet Size;

for(int i = 1; i <= n; i++)
{
        if(pI > 0)
        {
                dcI = dcI + qI;

                if(dcI > 0)
                {
                        do
                        {
                                // [Serve Packet]
                                dcI = dcI – pS;
                        }while(dcI >=0 && noP > 0);
                }
                if(dcI >= 0 && noP ==0) // for fairness issues
                {
                        dcI = 0; //  the dcI is reset
                }
        }
}
```

Table1: DRR scheduling algorithm

Usually the quantum is set to the maximum packet size. This is done in order to make sure that at least one packet per flow per non-empty queue is served.

### V. MODIFIED DEFICIT ROUND-ROBIN (MDRR)

MDRR scheduling is an extension of the previously mentioned DRR scheduling scheme. There may be different modifications of the DRR scheme and yet share the name is MDRR. Nevertheless, MDRR is mainly used as a scheduling scheme for the 12000 Cisco routers.

The algorithm depends on the DRR scheduling fundaments to a great extent, however, in MDRR the quantum value given to the queues is based on the weight associated with them, as indicated in Equation 1.

```
// q = Quantum;
// w = Weight;
// mtu = Maximum Tranmission Unit;

q = mtu + 512 * w;
```
Table 2: MDRR Equation (1).

Maximum Transmission Unit (MTU) is the maximum packet size that a queue may deliver. Note that, since the MTU is a constant number for a given system, quantum value and weight are therefore directly proportional and hereafter could be used interchangeably.

The reason of including the MTU parameter in equation 1 is to ensure that the quantum to be delivered to the intended queue at least enables the queue to transmit one packet. Since if no packet was transmitted in a round this results in the increase of the operational complexity. Except cases where the deficit counter is below zero.

In Equation 1 the weight is assigned to be equal to a percentage ratio and its indicated as shown in Equation (2).

```
// w = Weight;
// mtmr = Multi Transmit Multi Receive;
// sps = Symbol Per Second;
// tsc = Total System Capacity
w = mtmr(sps)/tsc(sps)*100;
```
Table 3: MDRR Equation (2).

The Cisco MDRR scheduling scheme adds a Priority Queue (PQ) into consideration with DRR. A Priority Queuing scheme isolates high demanding flows from the rest of the other flows for the reason of better quality of service provisioning. This is illustrated as shown in Figure 7.

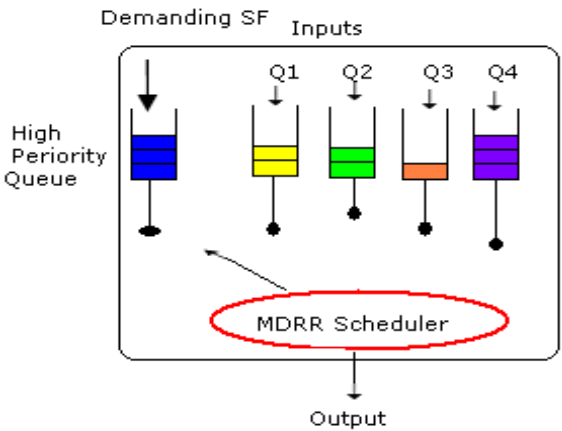
Figure 7: MDRR Scheduler

According to the mode of serving the Priority Queue, there are mainly two types of MDRR schemes:

- Alternate Mode: In this mode the high priority queue is serviced in between every other queue. For instance the scheduling sequence may be as follows: {PQ, Q1, PQ, Q2, PQ, Q3, PQ, and Q4}.

- Strict Priority Mode: here the high priority queue is served whenever there is backlog. After completely transmitting all its packets then the other queues are served. However, as soon as packets are backlogged again in the high priority queue, the scheduler transmits the packet currently being served and moves back to the high priority queue.

VI. MDRR ADJUSTMENTS

Priority queuing technique could be applied to classes of queues rather than the queues themselves individually. The intensive high-priority-queue scheduling idea instead could be achieved by assigning distinct and special weights to particular queues possessing large backlogs. For example, this could be shown as in Figure 8.

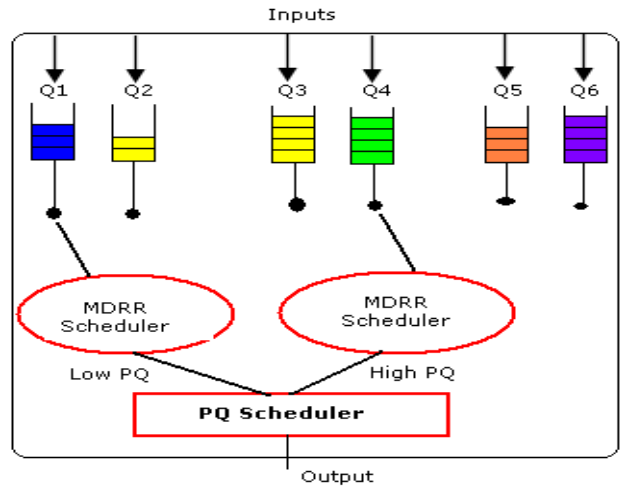
Figure 8: MDRR Scheduler adjustment

As indicated in Figure 8, the priority queuing scheduler gives precedence to the queues Q3, Q4, Q5 and Q6 which are scheduled by the high priority MDRR scheduler. After the high priority queues on the right hand side are completely served and no packet is found to be waiting in the buffers, then the PQ scheduler switches to the low priority class and serves the queues Q1 and Q2 in a MDRR scheduling scheme[6].

Changing the order of scheduling according to class priority, in other words, means to change the frequency of the scheduler serving a particular queue. This eventually leads to changing the throughput, latency, jitter and packet overflow at the queues. UGS and ertPS scheduling services are usually utilized for voice traffic and as it was mentioned before the requirements of thee scheduling services possess their own grant time period and polling time period, respectively. Thus to provide QoS, these strict constraints must be applied and there is not much scheduling flexibility. However, in this work it will be shown that rtPS may also carry voice traffic according to the voice related criteria. On the other hand, rtPS and nrtPS scheduling mechanisms are used for data traffic. In OPNET "Modeler Software – WiMAX Module"/Other Software MDRR can use for polling services (rtPS and nrtPS) while RR is used for BE services. Then priority queuing is applied between the two classes of queues. The high priority is given to the rtPS and nrtPS queues since they possess QoS constraints. Unlike rtPS and nrtPS, BE services do not have any QoS constraints and thus they are assigned as a low priority queue class[5].

Considering the case of using AMC in conjunction with the discussed MDRR scheduling scheme is the central idea of this paper. As it was mentioned before AMC adaptively changes the modulation and coding for the BS and MS according to the channel quality that the MS experiences. When a mobile is initiated communication with the BS while it is configured to operate in the AMC mode, it makes use of the most robust burst profile, which in the case of Mobile WiMAX it is QPSK ½. This makes sure that even if the MS is at its furthest distance from the BS, it is still capable of establishing initial communication. Since QPSK ½ mode is used then the Symbol Per Second (SPS) value of the MTMR is equal to the Bit Per Second (BPS) value of the MTMR. According to the Equation 1, having the lowest modulation burst profile leads to gaining the greatest weight. Thus in brief, all AMC operating mobile terminal start with the most robust modulation and coding profile and are assigned the highest weight [13-16].

This is determined to be as such because there is a probability that any mobile station may move away from the BS, then according to AMC, more robust modulation and coding shall be used. Consequently, this means only low bit rates will be available at the mobile terminal. So here, as a reimbursement, the mobile terminal is assigned more weight to compensate for its low data rate experience. Otherwise, it would be unfair to give distant mobile stations less data rate or bandwidth. However, the main point of this paper is that, mobile stations near the base station can actually have better data rates and without ignoring the fairness criteria. This idea originates from the fact that mobile stations near the BS are changing their modulation format and coding rate as soon as they establish communication with the BS. The change in modulation and coding scheme will be towards a higher modulation format and a lower coding rate. This is valid, since mobile stations near the base station experience very good channel impulse responses. As a result, these mobile stations consume less bandwidth than the distant mobile stations. Thus, in order not to waste the bandwidth allocated to the mobile stations near the base station, it is suggested that the more weight should be given to mobile stations near the BS.

In Mobile WiMAX, the channel called Channel Quality Indicator Channel (CQICH) is dedicated to return channel quality measurements, measured by the SS, to the BS. The parameter sent over this channel is the CINR value. For the purpose of adjusting the weight of the channel, CINR will be incorporated into the design as indicated in Equation (3) below.

```
// w = Weight;
// mtmr = Multi Transmit Multi Receive;
// sps = Symbol Per Second;
// tsc = Total System Capacity;
//cinr = Carrier to Interference and Noise Ratio;
//int cinrInt;

cinrInt = (cinr-12/22) * 3.5;
w = (mtmr(sps)/tsc(sps) * 100) + cinrInt * 3;
```
Table 4: MDRR Equation (3).

The numbers associated with the CINR portion of Equation (3) is designed after intensive experimentation. Basically, the right hand portion is an additional amount of weight given to those mobile stations with considerable CINR values. The CINR values noticed from experiment results range from 15 dB to 30 dB. Equation (3) will be incorporated into the OPNET MDRR scheduler and then results will be examined with taking the CINR part into account in implementation work using software.

VII. SYSTEM MODEL AND SCENARIOS

*A. Evaluation Methodology*

There are three main methods to evaluate a Scheduling algorithm: **Deterministic modeling:** Different algorithms are tested against a predetermined workload. Then the performance results are compared. **Queuing models:** Random backlogs are studied by analytically – in a mathematical way. **Implementation/Simulation:** The most versatile method of testing scheduling algorithms is to actually simulate the designed algorithm with real life data and conditions.

*B. OPNET Modeler*

Indeed trying to analyze and simulate a part of the total system would still need the fully simulated and accurate system beforehand. This is supported by the fact that, any other cooperating system entity may totally or in part change the outcome of a simple scenario. Therefore, as many nodes and sub-nodes, with their correct attributes, should be incorporated into the system model. However, this is not an easy task at all. Even a very small portion of the whole system model would need weeks to get it to behave like the original real life system. It should also be mentioned that, lacking a fully simulated system model, usually causes researchers to fail at some point during their study to get feasible results.

It is true that commonly used soft wares like C, C++, Java, MATLAB and many other programming languages are strong and performing languages; however, these programmes do not come with a model of a specific system. Thus for the sake of accuracy accompanied by an almost complete WiMAX system model simulation OPNET Modeler was researched to be one of the best candidates. OPNET Modeler is a product of the OPNET Technologies Inc. It is a Discrete Event Simulation (DES) programme: events are handled in a chronological manner. It has a Graphical User Interface (GUI) with a "user friendly" sense. It has an enormous library at the service of the user. On request, OPNET Technologies can provide ready-to-use models. For the research to be done in this project, "OPNET Modeller Wireless Suite" was provided with an accompanying "WiMAX Model"[16].

OPNET Modeller is made up of 4 main editors: **Project editor:** This is where the wireless network is laid out. The network comprises of the network nodes and connections. **Node editor:** Nodes can be edited with the node editor. Nodes are made up of sub-modules, which carry out the functions of the node. The functions are in the form of processes. **Process editor:** Processes function as a Finite State Machine (FSM). **Code editor:** Each state in the FSM is coded in either C or C++ or Java programming language. The code is actually the tasks that the node does.

*C. System Mode*

In the following scenarios 6 mobile stations are configured in operate in the AMC mode. The load on the network mobile stations is 96 Kbps, originally 64Kbps, but with header and control information it adds up to 96 Kbps. The parameters of the mobile stations, Base station, WiMAX Configuration Node are as shown in Figure 8, Figure 9, Figure 10, and Figure 11.

Figure 8: WiMAX Mobile Station Parameters

Figure 9: WiMAX Base Station Parameters

Figure 10: WiMAX Configuration Node Parameters

Figure 11: WiMAX Configuration Node Parameters

*D. Simulated Scenario*

In the following scenarios the parameters for the Mobile Station, Base Station and the WiMAX configuration nodes

were set as they were previously shown in Figures 8, 9, 10 and 11.

In this scenario equation (3) was set into the MDRR scheduling C code and the following graphs were obtained after simulating for 1 minute and 40 seconds. Figure 12 shows the CINR gotten as a feed back from the mobile station to the base station.

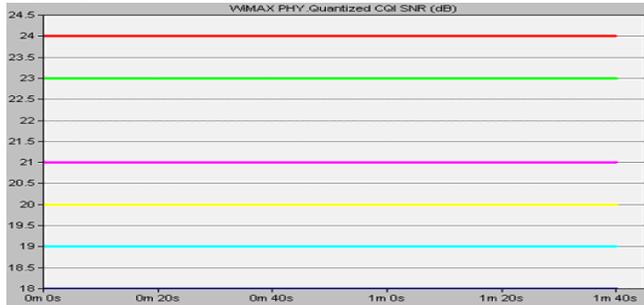
Figure 12: Mobile Station CINR Values

Figure 12 indicates that, MS_0 has the highest CINR value and MS has the lowest CINR value.

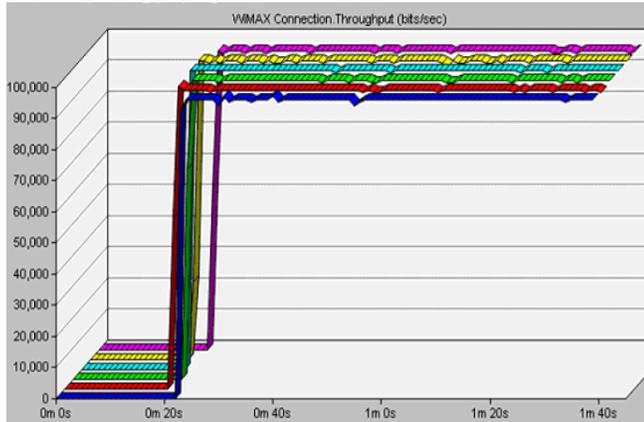
Figure 13: WiMAX UL MS Throughput

It is shown in Figure 13 that all the Mobile station UL connections do transmit the work load applied on them. This may indicate that the weights assigned to the queues were affecting the dequeuing process enough to eventually lower the throughputs of the corresponding mobile stations.

On the other hand, polling the queues might have had more effect on the throughput rather than the assigned weights. Since however the weight may be still polling the SS has a superior act on servicing that SS. This was justified when the MTMR (which polling period is dependent on) of MS_2 was set to be higher than the other queue. The result, as shown in Figure 14, indicate that MTMR indeed have raised the throughput of the corresponding queue[16].

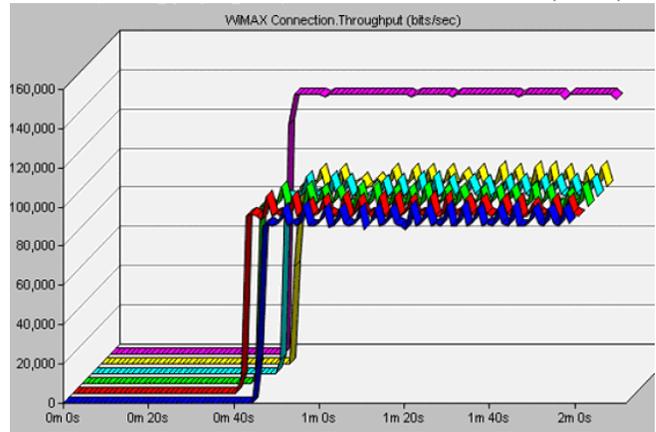
Figure 14: The effect of MTMR on UL throughput

Considering delays that the mobile stations possess, Figure 15 shows the how much delay each mobile station is experiencing.

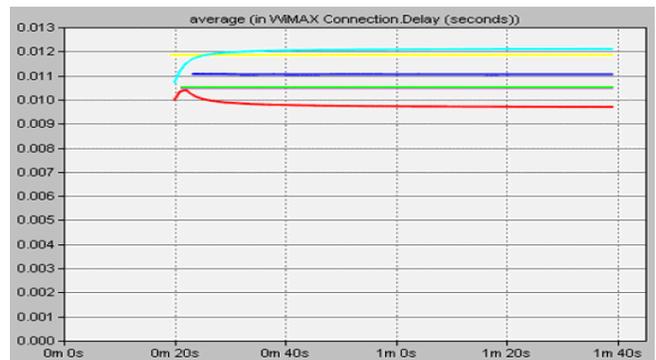
Figure 15: Mobile Station Delay

As was expected Mobile station 4,1 and 0 (green, pink and red) are at the bottom, meaning that they own the lowest delays. This may have been the result of the weighting process. In other words, when great weights were assigned to mobile stations with good channel quality, then the MDRR scheme have served them more than the others; Causing the other queues to wait till they get some deficit. Moreover, the Figure also indicates that on average very low delays have been achieved. This is a very good indication, which might qualify the scheduler to be utilized for voice applications as well. Nevertheless, further simulations are required to approve that the algorithm is capable of handling large amounts of users.

*E. Resulted Simulation Limitations and Constraints*

OPNET Modeler Software is still immature. There are many bugs in the WiMAX Model which took a lot of time of study. Furthermore, the software is quite complicated, requiring that the user is already a professional C programmer to be able to change the code within, which eventually blocked the way to give complete simulation set. Some of the basic mistakes were encountered regarding the position of the mobile stations. For instance, a mobile station at a distance 1 Km from the BS in one direction does not have the CINR as in another direction, knowing that the antenna is an omni antenna and the position should be a problem if the distance was is same.

Figure 16: An OPNET bug

It is noticed here (Yellow & Orange color), when the Mobile Station "MS" moves, it gets dropped off exactly right after that the software produces an error, which should not happen. This means that the software itself also has some bugs yet to be corrected. Eventually, after re-simulating the programme a couple of times, the bug unexpectedly disappeared. Many other bugs were noticed and took a lot of time for the software to get back to normal. Another simulation limitation as that Modeler sets the path taken by the uplink to be the same as the path taken by the downlink for CINR measurements, where this is almost always not the case, but rather the paths take different channel quality values.

CONCLUSION

Being one of the hottest current research issues and as indicated by Samsung: designing a scheduler that is less complex, more efficient and provides a superior Quality of Service is of great importance to mobile WiMAX systems. In this paper, a comprehensive, yet brief, introduction was given for the IEEE 802.16e commercially known as Mobile WiMAX Systems. Modified DRR scheduling algorithm has been studied in depth and implemented in OPNET Modeler Software. After that it was attempted to enhance the throughput of the system with regard to the channel quality if the subscriber stations while taking fairness into consideration.

AUTHORS PROFILE

**C. Ravichandiran** received the MCA from the Madurai Kamaraj University, India, in 1999. He received the M. Tech degree in Software Engineering from IASE University, India. And currently pursuing PhD degree in Computer Science from SASTRA University, India. His fields of interest are Computer Networks, Network Security, Wireless and Mobile Communication, Database. He has more than 9 publications to his credit in international journals and conferences. He is a life member of the International Association of Computer Science and Information Technology (IACSIT), International Journal of Electronic Business (IJEB), and International Association of Engineers (IAENG).

**Dr. C. Pethuru Raj** (www.peterindia.net) has been working as a lead architect in the corporate research (CR) division of Robert Bosch. The previous assignment was with Wipro Technologies as senior consultant and was focusing on some of the strategic technologies such as SOA, EDA, and Cloud Computing for three years. Before that, he worked in a couple of research assignments in leading Japanese universities for 3.5 years. He has 8 years of IT industry experiences after the successful completion of his UGC-sponsored PhD in formal language theory / fine automata in the year 1997. He worked as a CSIR research associate in the department of computer science and automation (CSA), Indian institute of science (IISc), Bangalore for 14 memorable months. He has been authoring research papers for leading journals and is currently involved in writing a comprehensive and informative book on Next-Generation Service Oriented Architecture (SOA).

**Dr. V. Vaidhyanathan** received the PhD degree from the Alagappa University, Karaikudi, India. He is currently Professor and HOD-IT in School of Computing, SASTRA University, and Thanjavur, India. He has more than 19 years of experience in teaching and research. He has been guiding more than 25 M.Tech Projects, 5 PhD and thesis. His fields of interests are various techniques in Image Processing, Computer vision for shape identification, Reconstruction, noise removal, online correction of an image by developing software and in the area of cryptography. Various applications of soft computing techniques for object identifications. He has more than 40 publications to his credit in international journals and conferences. He has visited many universities in India.